\documentclass[twocolumn,footinbib,aps,pra,10pt]{revtex4-2}
\usepackage{amssymb}
\usepackage{graphicx}
\usepackage{amsmath}
\usepackage{color}
\usepackage{xcolor}
\usepackage{subfigure}
\usepackage{mathrsfs}
\usepackage{bm}
\usepackage{cancel}
\usepackage{braket}
\usepackage[normalem]{ulem}

\usepackage{times}
\usepackage{amsmath}
\usepackage{amsfonts}
\usepackage{amssymb}
\usepackage{bm}
\usepackage{graphicx}
\usepackage{epstopdf}
\usepackage{enumerate}
\usepackage{color}

\graphicspath{{img/}}

\begin{document}
\title{Quantum Gates Robust to Secular Amplitude Drifts}
\author{Qile David Su}
\email[Contact at (he/him/his): ]{qilesu@g.ucla.edu}
\author{Robijn Bruinsma}
\author{Wesley C. Campbell}
\affiliation{Department of Physics and Astronomy, University of California, Los Angeles, CA 90095, USA}

\begin{abstract}
Quantum gates are typically vulnerable to imperfections in the classical control fields applied to physical qubits to drive the gates. One approach to reduce this source of error is to break the gate into parts, known as \textit{composite pulses} (CPs), that typically leverage the constancy of the error over time to mitigate its impact on gate fidelity. Here we extend this technique to suppress \textit{secular drifts} in Rabi frequency by regarding them as sums of \textit{power-law drifts} whose first-order effects on over- or under-rotation of the state vector add linearly. Power-law drifts have the form $t^p$ where $t$ is time and the constant $p$ is its power. We show that composite pulses that suppress all power-law drifts with $p \leq n$ are also high-pass filters of \textit{filter order} $n+1$ \cite{ball_walsh-synthesized_2015}. We present sequences that satisfy our proposed \textit{power-law amplitude} criteria, $\text{PLA}(n)$, obtained with this technique, and compare their simulated performance under time-dependent amplitude errors to some traditional composite pulse sequences. We find that there is a range of noise frequencies for which the $\text{PLA}(n)$ sequences provide more error suppression than the traditional sequences, but in the low frequency limit, non-linear effects become more important for gate fidelity than frequency roll-off. As a result, the previously known $F_1$ sequence, which is one of the two solutions to the $\text{PLA}(1)$ criteria and furnishes suppression of both linear secular drift and the first order nonlinear effects, is a sharper noise filter than any of the other $\text{PLA}(n)$ sequences in the low frequency limit.
\end{abstract}
\maketitle

\section{Introduction}
\label{sec.introduction}

In the network model, quantum computation (QC) relies on applying quantum logic gates to qubits. In physical QC implementations, the gates are typically realized with applied classical fields (hereafter, \emph{pulses}) that are necessarily imperfect \cite{van_dijk_impact_2019, ball_role_2016, vandersypen_nmr_2005}. Unwanted variations in the power and duration of the control pulses, for example, due to heating of the amplifier, can cause over- or under- rotations of the state vector, which we will refer to as amplitude errors. Alternatively, instability of the frequency of the pulses (or, equivalently, the energy splitting of the qubit states) causes the control pulses to stray from resonance, which we will refer to detuning errors. These imperfections are a major obstacle to QC \cite{brown_single-qubit-gate_2011}. While quantum error correction provides a way to perform fault tolerant calculation in principle, the error probability per gate must still be reduced to less than a certain critical value, estimates for which go as low as $10^{-6}$ \cite{knill_quantum_2010, gottesman_stabilizer_1997, knill_quantum_2005, knill_resilient_1998}.

The general question of what control pulses are necessary to best achieve some target gate under the effect of uncontrollable factors is answered by quantum optimum control \cite{glaser_training_2015}. It is further divided into open-loop and closed-loop protocols depending on whether the computation of the optimal control involves experimental feedback. Both analytical and numerical optimization are employed to produce the optimal control parameters. In the specific scenario we study, since the characteristic timescale of pulse imperfections is typically long compared to the duration of the pulse, strategies to mitigate static errors (effectively, by depolarizing the error using techniques similar to spin echo) have been employed extensively in QC. It has been shown that infidelity from static amplitude and/or detuning errors can be suppressed using composite pulses \cite{brown_arbitrarily_2004, merrill_progress_2014, low_optimal_2014, levitt_composite_1986, husain_further_2013, odedra_dual-compensated_2012, wimperis_broadband_1994, husain_further_2013-1, torosov_composite_2019, bando_concatenated_2013, souza_experimental_2012}. Dynamically corrected gates can reduce the error per gate in open quantum systems undergoing
linear decoherence \cite{khodjasteh_dynamically_2009}. The principle behind both techniques is that by careful construction of the pulse sequence (or simply, sequence), errors accumulated earlier in a sequence are cancelled by those acquired later. Certain pulse sequences, such as the Knill sequence (5 successive $\pi$ pulses with phases 30$^\circ$, 0$^\circ$, 90$^\circ$, 0$^\circ$, 30$^\circ$) \cite{ryan_robust_2010, jones_designing_2013} and the $F_1$ sequence (5 successive $\pi$ pulses with phases 46.6$^\circ$, 255.5$^\circ$, 0$^\circ$, 104.5$^\circ$, 313.4$^\circ$) \cite{wimperis_iterative_1991}, cancel the error contributed by the leading order term in the Magnus expansion \cite{magnus_exponential_1954} of the time evolution operator. The Knill sequence simultaneously suppresses static amplitude and detuning errors to first order in the Magnus expansion, while the $F_1$ sequence suppresses static amplitude errors to second order.

It has been verified both numerically and experimentally that pulse sequences designed for static errors can, to some extent, suppress errors at the low frequency end of the power spectral density \cite{kabytayev_robustness_2014, soare_experimental_2014, zhen_experimental_2016}. A key concept is the pulse sequence's \textit{filter function}, analogous to the transfer function for electrical filters, which characterizes the frequency domain response of the infidelity of the operation achieved by the pulse sequence to the time-dependent drift experienced by the system \cite{green_arbitrary_2013, cywinski_how_2008, biercuk_dynamical_2011, ball_walsh-synthesized_2015, paz-silva_general_2014, kabytayev_quantum_2015}. Research in dynamical decoupling, which uses composite pulses to maintain quantum memory, has found that the behavior of the filter function near zero frequency, as determined by its Taylor expansion, is an important characteristic of the pulse sequence \cite{cywinski_how_2008, ball_walsh-synthesized_2015, uhrig_keeping_2007, uhrig_exact_2008}. Expansions of drifts in the time domain has also been used as a criteria for sequence design under dephasing noise and in dynamical decoupling \cite{szwer_keeping_2010}. A significant difference in objective between dynamical decoupling and the design of robust gates is that, for the latter, we do not impose a requirement on the sequence duration. Rather we seek the shortest possible robust sequences given limited control strengths. This negatively impacts sequences with more pulses, since they are more time-consuming and experience control imperfection for a longer duration.

In this paper, we use composite pulses to construct, for the subset of logical operations corresponding to $\pi$-rotations of a single qubit about a transverse axis, robust gates that suppress drifts in the amplitude of the control pulses, specifically, \textit{secular drifts}, \textit{i.e.}, those that occur at frequencies smaller than the Rabi frequency (proportional to the amplitude) of the pulses. We assume that the drifts are analytic in time, thus admitting Taylor expansions into \textit{power-law drifts} whose time dependence has the form $t^n$ where $t$ is time and $n$ is the power. Since we intuitively expect the lowest powers of time to be the most important, we will design composite pulses that successively cancel the effect from each term in the Taylor expansion of the drift. It turns out that a pulse sequence’s response to power-law drifts is closely connected to the roll-off (the slope on a log-log plot) of its filter function near zero frequency, sometimes used to define the \textit{filter order} \cite{ball_walsh-synthesized_2015}. While the design criteria we introduce do not depend on the frequency domain description, this close connection provides evidence that these pulse sequences are high-pass filters. We illustrate this by calculating the filter function analytically and running numerical simulations.

The rest of this paper is organized as follows. Section \ref{sec.background} defines quantities relevant to the problem and expresses the gate infidelity with terms in the Magnus expansion. Section \ref{sec.secular-drifts} Taylor expands a secular drift in the pulse amplitude into power-law drifts and proposes the power-law amplitude $\text{PLA}(n)$ criteria, which constrains the pulse sequence. Section \ref{sec.random-errors} relates suppression of power-law drifts to the low frequency roll-off of the pulse sequence. In section \ref{sec.analytically} we solve the $\text{PLA}(1)$ constraints analytically and show that the $F_1$ sequence is in fact one of its two distinct solutions, the other of which we label $[\text{PLA}(1)]_2$. In section \ref{sec.numerically} we solve the $\text{PLA}(n)$ constraints for $n > 1$ numerically. In section \ref{sec.beyond-linear-response}, we numerically simulate the frequency response of the sequences that satisfy the $\text{PLA}(n)$ criteria, and compare their performance to a few known pulse sequences.

\section{Theory}
\label{sec.theory}
\subsection{Background}
\label{sec.background}
We consider a single, ideal qubit consisting of an isolated, stable, two-level system in which transitions between the two states can be induced by oscillating control fields resonant with the qubit splitting \cite{kabytayev_robustness_2014, green_arbitrary_2013} and focus on its behavior in the presence of non-ideal control fields. A pulse sequence is made from $N$ simple pulses that are rotations of the qubit's state vector (Bloch vector) around single axes. Here we assume the ideal pulses have square envelopes, the rotation axes are always confined to the equatorial plane of the Bloch sphere, and there is no delay time in between pulses. Generalization to non-square pulses without significant changes to our conclusions can be found in appendix \ref{sec.non-square-pulses}. The $l$th pulse begins at time $t_{l-1}$ and ends at time $t_{l}$. The error-free control-field amplitude of pulse $l$ (which we will identify as the Rabi frequency) is $\Omega_l$, and its phase (which sets the rotation axis) is $\phi_l$. Finally, we denote $\tau \equiv t_N$. In the rotating frame, after suitable approximations, the error-free Hamiltonian ($\hbar \equiv 1$) is given by
\begin{equation}
	H_0(t) = \sum_{l=1}^NG^{(l)}(t)\frac{\Omega_l}{2}\boldsymbol{\rho}_a^{(l)}\cdot\boldsymbol{\hat\sigma}, \label{eq.hamiltonian-error-free}
\end{equation}
where we define the vectors $\boldsymbol{\rho}_a^{(l)} \equiv \boldsymbol\rho(\phi_l)\equiv(\cos\phi_l,\sin\phi_l,0)$, $\boldsymbol{\hat\sigma} \equiv (\hat{\sigma}_x, \hat{\sigma}_y,\hat{\sigma}_z)$, and the square envelope function is defined as $G^{(l)}(t) \equiv 1$ for $t_{l-1}< t <t_l$ and 0 otherwise. We denote $U_0(t)$ as the unitary time evolution operator generated by $H_0(t)$ (\textit{i.e.}, when errors are zero).

We model pulse imperfections by $\beta_a(t)$ and $\beta_d(t)$ (where $t$ is time) for the amplitude and frequency errors, respectively. The full Hamiltonian is
\begin{align}
	H(t) &= \sum_{l=1}^NG^{(l)}(t)\frac{[\Omega_l+\beta_a(t)]}{2}\boldsymbol{\rho}_a^{(l)}\cdot\boldsymbol{\hat\sigma} + \frac{\beta_d(t)}{2}\sigma_z \\
	&= H_0(t) + \nonumber\\
	&\quad \left[\beta_a(t)\left(\frac{1}{2}\sum_{l=1}^NG^{(l)}(t)\boldsymbol{\rho}_a^{(l)}\right) + \beta_d(t)\left(\frac{1}{2}\mathbf{\hat{z}}\right)\right]\cdot\boldsymbol{\hat\sigma} \\
	&= H_0(t) + H_{\mathrm{err}}(t). \label{eq.hamiltonian}
\end{align}
For unitary operations, the operational fidelity is
\cite{kabytayev_robustness_2014, green_arbitrary_2013},
\begin{equation}
	\mathcal{F} \equiv \frac{1}{4}\left|\mathrm{Tr}\left[U_0^{\dagger}(\tau) U(\tau)\right]\right|^2, \label{eq.fidelity}
\end{equation}
where the trace term is the Hilbert-Schmidt or trace inner product \cite{nielsen_quantum_2010} between the ideal ($U_0(\tau)$) and actual ($U(\tau)$) evolution operators.

To separate the evolution from $H_0(t)$ and $H_{\text{err}}(t)$, we work in the \emph{toggling frame} \cite{suter_recursive_1987, odedra_dual-compensated_2012, green_arbitrary_2013} (the interaction frame with respect to the errorless pulse sequence \cite{levitt_composite_1986}) where the effective Hamiltonian is $H'(t) = U_0^{\dagger}(t)H_{\mathrm{err}}(t)U_0(t)$. If we perform the Magnus expansion of the time evolution operator in the toggling frame \cite{green_arbitrary_2013} and then transform back to the frame co-rotating with the controls, the full evolution operator is the product of the error-free evolution operator and a correction due to $H_{\text{err}}(t)$,
\begin{equation}
	U(\tau) = U_0(\tau)\exp\left[-\mathrm{i}\mathbf{a}\cdot\boldsymbol{\hat{\sigma}}\right] \label{toggling-frame-transformation}
\end{equation}
where $\mathbf{a} = \mathbf{a}_1 + \mathbf{a}_2 + \cdots$,
\begin{align}
	\mathbf{a}_1 &= \int_0^{\tau}\mathrm{d}t_1 \left[\beta_a(t_1)\boldsymbol{\rho}_a(t_1) + \beta_d(t_1)\boldsymbol{\rho}_d(t_1)\right] \label{eq.magnus-expansion-first-order}\\
	\mathbf{a}_2 &= \int_0^{\tau}\mathrm{d}t_1\int_0^{t_1}\mathrm{d}t_2 \left[\beta_a(t_1)\boldsymbol{\rho}_a(t_1) + \beta_d(t_1)\boldsymbol{\rho}_d(t_1)\right]\times \nonumber \\
	&\quad\quad\left[\beta_a(t_2)\boldsymbol{\rho}_a(t_2) + \beta_d(t_2)\boldsymbol{\rho}_d(t_2)\right] \label{eq.magnus-expansion-second-order}\\
	&\cdots \nonumber
\end{align}
$\boldsymbol{\rho}_a(t)$ and $\boldsymbol{\rho}_d(t)$ (``control vectors") are one-half the unit vectors along the error rotation axes in the toggling frame and depends on the parameters of the control pulse \cite{green_arbitrary_2013}.
The infideility can be written in terms of $\mathbf{a}_i$,
\begin{align}
	1 - \mathcal{F} &= 1 - \frac{1}{4}\left|\mathrm{Tr}[\exp(-\mathrm{i}\mathbf{a}\cdot\boldsymbol{\hat\sigma})]\right|^2 = 1 - \cos^2(|\mathbf{a}|) \\
	&= |\mathbf{a}|^2 - \frac{1}{3} |\mathbf{a}|^4 + \cdots \\
	&= (\mathbf{a}_1 + \mathbf{a}_2 +\mathbf{a}_3 + \cdots)^2 \nonumber \\
	&\quad - \frac{1}{3}\left[(\mathbf{a}_1 + \mathbf{a}_2 +\mathbf{a}_3 + \cdots)^2\right]^2 + \cdots \\
    &= |\mathbf{a}_1|^2 + \left(2 \mathbf{a}_1\cdot\mathbf{a}_2\right) + \left(\mathbf{a}_2^2 -\frac{1}{3}|\mathbf{a}_1|^4 + 2 \mathbf{a}_1 \cdot \mathbf{a}_3\right) \nonumber \\
    &\quad + \cdots. \label{eq.infidelity}
\end{align}
By zeroing $\mathbf{a}_1$, infidelity can be suppressed to first order in the magnitudes of $\beta_a(t)$ and $\beta_d(t)$. Our analysis within this section will be done in the linear response regime where $\mathbf{a}_1$ dominates, so we may write
\begin{equation}
	1 - \mathcal{F} \approx |\mathbf{a}_1|^2. \label{eq.infidelity-approx}
\end{equation}
In addition, from here onward we will set $\beta_d(t) = 0$, restricting ourselves to amplitude errors only. The errorless Rabi frequency $\Omega$ will be a constant.

\subsection{Secular drifts}
\label{sec.secular-drifts}
In real devices, time-dependent imperfections in pulse amplitude (or, more generally, pulse area) can occur, for instance due to the temperatures of components changing during uneven use \cite{blume-kohout_demonstration_2017, brown_single-qubit-gate_2011}.  To model this, we let $\beta_a(t)$ describe a secular drift, \textit{i.e.}, a drift whose characteristic timescale is much longer than a pulse. We assume that $\beta_a(t)$ (where $t$ is time) is analytic, so it can be expressed as a power series in time,
\begin{equation}
	\beta_a(t) = \sum_{p = 0}^{\infty} \frac{1}{p!}\beta_a^{(p)}t^p, \label{eq.power-series-error}
\end{equation}
where $\beta_a^{(p)}$ is the $p$-th time derivative of $\beta_a(t)$ evaluated at $t=0$. Plugging this into Eq.\eqref{eq.magnus-expansion-first-order}, we get that the first order correction $\mathbf{a}_1$ is
\begin{align}
	\mathbf{a}_1 = \sum_{p=0}^{\infty}\beta_a^{(p)}\mathbf{c}_{a, p},
\end{align}
where
\begin{equation}
	\mathbf{c}_{a, p} \equiv \frac{1}{p!}\int_0^{\tau}t^p\boldsymbol{\rho}_a(t)\mathrm{d}t. \label{eq.constraint-power-law-drift-lhs}
\end{equation}
To suppress a secular drift up to and including a power-law $t^n$ dependence, the task is to design $\boldsymbol{\rho}_a(t)$ such that the following is satisfied:
\begin{equation}
	\mathbf{c}_{a, p} = 0\quad \text{for } p = 0, 1, \cdots, n.  \label{eq.constraint-power-law-drift}
\end{equation}
When this is achieved, the infidelity is, to leading order,
\begin{equation}
	1 - \mathcal{F} \approx |\mathbf{a}_1|^2 \approx \left(\beta_a^{(n + 1)}\right)^2(\mathbf{c}_{a, n + 1})^2. \label{eq.leading-order-infidelity}
\end{equation}


Now we write Eq.\eqref{eq.constraint-power-law-drift} in terms of parameters of a composite pulse, which requires knowledge of its possible forms. In general, the net effect of composite pulse changes wildly when the parameters of each of the constituent pulses are allowed to vary. For this reason, we choose to focus on pulse sequences that perform a net $\pi$-pulse around the $x$-axis (an $\hat{X}$ gate) in this work. Such a choice (as opposed to, say, a $\pi/2$- or $\pi/4$-pulse) leads to simplified sequences because the target gate can always be constructed from an odd number of $\pi$-pulses whose phases can be freely varied without compromising the target $\hat{X}$ gate. The form for the composite pulse sequence will be 
\begin{equation}
	\pi_{\phi_1}\to\pi_{\phi_2}\to\cdots\to\pi_{\phi_{N}}\quad\ (N\ \text{is odd}), \label{eq.CP-form-pi}
\end{equation}
where the notation $\pi_{\phi_l}$ means a $\pi$-pulse with phase $\phi_l$, and $\phi_l$ are constrained by the following condition to ensure an $\hat{X}$ gate,
\begin{equation}
	g(\boldsymbol{\phi}) \equiv \sum_{l=1}^N (-1)^l\phi_l = k\pi\quad \text{for integer } k. \label{eq.overall-constraint}
\end{equation}
For a gate such as $(\cos\gamma\ \hat{X} + \sin\gamma\ \hat{Y})$, add $\gamma$ to the right hand side of Eq.\eqref{eq.overall-constraint}. It has been shown that robust sequences for rotations other than $\pi$ can in some cases be constructed by prepending or appending a sequence of $\pi$-pulses to a target gate \cite{brown_single-qubit-gate_2011, low_optimal_2014}.

Under the composite pulse assumed in Eq.\eqref{eq.CP-form-pi}, we get $t_{l-1} = (l-1)\pi/\Omega$. Since $\boldsymbol{\rho}_a$ is one-half the unit vector along the error rotation axes in the toggling frame, we can rewrite Eq.\eqref{eq.constraint-power-law-drift-lhs} as
\begin{align}
	\mathbf{c}_{a, p} &= \frac{1}{2p!}\sum_{l=1}^N\boldsymbol{\rho}(\phi_l')\int_{t_{l-1}}^{t_l}t^p\mathrm{d}t
\end{align}
where $\phi_j'$ are the \textit{toggling-frame phases} \cite{wimperis_broadband_1994},
\begin{equation}
	\phi_j' \equiv -(-1)^j\phi_j-\sum_{k=1}^{j-1}(-1)^k2\phi_k, \label{eq.toggling-frame-phases}
\end{equation}
and as before $\boldsymbol\rho(\phi_l') \equiv (\cos\phi_l', \sin\phi_l', 0)$. Integrating and expanding with the binomial theorem gives us
\begin{align}
	\mathbf{c}_{a, p} &= \frac{1}{2p!}\sum_{l=1}^N\boldsymbol{\rho}(\phi_l')\int_{t_{l-1}}^{t_l}t^p\mathrm{d}t \\
	&= \frac{1}{2}\left(\frac{\pi}{\Omega}\right)^{p+1}\frac{1}{(p+1)!}\sum_{l=1}^N\boldsymbol{\rho}(\phi_l')\left[l^{p+1} - (l-1)^{p+1}\right] \\
	&= \frac{1}{2}\left(\frac{\pi}{\Omega}\right)^{p+1}\sum_{q=0}^p\frac{1}{q!(p-q+1)!} \mathbf{c}'_{a, q},
\end{align}
where we define
\begin{equation}
	\mathbf{c}'_{a, p} \equiv \sum_{l=1}^N(l-1)^p\boldsymbol{\rho}(\phi_l'). \label{eq.constraint-discrete-drift-lhs-sc-seqs}
\end{equation}
Therefore the constraints in Eq.\eqref{eq.constraint-power-law-drift} are equivalent to
\begin{equation}
	\mathbf{c}'_{a, p} = 0\quad \text{for } p = 0, 1, \cdots, n. \label{eq.constraint-discrete-drift}
\end{equation}
This says that in the toggling frame the error rotation axes weighted by the $p$th power of the pulse number must sum to zero. We label Eq.\eqref{eq.constraint-discrete-drift} and Eq.\eqref{eq.overall-constraint} as the power-law amplitude $\text{PLA}(n)$ criteria. They will be solved in section \ref{sec.solving-the-constraints}. Solutions will be labeled $[\text{PLA}(n)]_i$, where $i$ is an integer index which differentiates distinct solutions.

Although by our construction, we suppress more power-law drifts within the linear regime as $n$ increases, there are also drawbacks from having a large $n$. First, the number of pulses (degrees of freedom) necessary to satisfy the constraints also increases. Since the Rabi frequency cannot be arbitrarily increased in practice, the duration of the sequence must increase with the number of pulses. There is another problem with long sequences within our framework that arises from unsuppressed residual power-law drifts. From Eq.\eqref{eq.leading-order-infidelity}, the infidelity of sequences that satisfy the $\text{PLA}(n)$ criteria are affected by the $(n+1)$th order power-law drift through $\mathbf{c}_{a, n+1}$, defined in Eq.\eqref{eq.constraint-power-law-drift-lhs}. Roughly speaking, since the duration of the sequence depend on $\tau$,
\begin{equation}
    |\mathbf{c}_{a, n+1}|\sim \frac{1}{(n+1)!}\int_0^{\tau(n)}t^{n+1} \mathrm{d}t = \frac{[\tau(n)]^{n+2}}{(n+2)!}.
\end{equation}
For a simple estimate, suppose that $\beta_a(t) = A\cos(\omega t)$, which models some sinusoidal modulation of the Rabi frequency. Then its time derivatives are $\left|\beta_a^{(n+1)}\right| = A\omega^{n+1}$. We can estimate the infidelity as
\begin{align}
    1 - \mathcal{F} &\sim A\frac{\omega^{n+1}[\tau(n)]^{n+2}}{(n+2)!}.
\end{align}
If $\tau$ increases fast enough with $n$, and if $\omega\tau$ is sufficiently large, then infidelity can potentially worsen as $n$ increases. These issues put a practical (so far unknown) limit on $n$ for $\text{PLA}(n)$ sequences, and we should examine them in a case-by-case basis. Once we solve the constraints and find some sequences (section \ref{sec.solving-the-constraints}), we will know exactly how much longer the sequences need to be; if we simulate the performance of these sequences (section \ref{sec.beyond-linear-response}), we will be able to determine unambiguously how much infidelity is affected by the sequence duration.

\subsection{Random errors}
\label{sec.random-errors}
Following \cite{green_arbitrary_2013, kabytayev_robustness_2014}, suppose the amplitude error $\beta_a(t)$ is a stationary process, has zero mean, and admits auto-covariance function defined by $R_a(t) = \langle\beta_a(t_0)\beta_a(t_0 + t)\rangle$. The power spectral density of $\beta_a(t)$ is
\begin{equation}
	S_a(f) \equiv \int_{-\infty}^{\infty}R_a(t)e^{-\mathrm{i}2\pi ft}\mathrm{d}t. \label{eq.power-spectral-density}
\end{equation}
In practice, the fidelity is measured by repeating the experiment, so we average Eq.\eqref{eq.infidelity} over an ensemble of $\beta_a(t)$ (the noise ensemble). Using Eq.\eqref{eq.magnus-expansion-first-order} to rewrite the dominant term in the infidelity (Eq. \eqref{eq.infidelity-approx}) with frequency domain variables, we get
\begin{equation}
	 1 - \langle\mathcal{F}\rangle \approx \langle |\mathbf{a}_1|^2\rangle = \int_{-\infty}^{\infty}\mathrm{d}f\frac{1}{(2\pi f)^2}S_a(f)h_a(f), \label{eq.infidel-from-filter-function}
\end{equation}
where $h_a(f)$, the \textit{filter function} \cite{green_arbitrary_2013, kabytayev_robustness_2014},
is given by
\begin{equation}
	h_a(f) \equiv \tilde{\boldsymbol{\rho}}_a^*(f)\cdot\tilde{\boldsymbol{\rho}}_a(f), \label{eq.filter-function}
\end{equation}
and
\begin{equation}
	\boldsymbol{\tilde{\rho}}_a(f) \equiv -\mathrm{i}2\pi f\int_0^{\tau}\boldsymbol{\rho}_a(t)e^{\mathrm{i}2\pi ft}\mathrm{d}t. \label{eq.control-vector-FT}
\end{equation}
To suppress low frequency errors, the filter function should have a steep roll-off in the low frequency limit. This is captured by its Taylor series expansion at $f = 0$. Appendix \ref{sec.filter-function-low-freq} shows that, if Eq.\eqref{eq.constraint-power-law-drift} is satisfied, then the leading order term in this expansion is
\begin{equation}
    h_a(f) \approx \left(\mathbf{c}_{a, n + 1}\right)^2(2\pi f)^{2n + 4}. \label{eq.low-freq-error-conclusion}
\end{equation}
Thus, if a pulse sequence suppresses power-law drifts, it automatically suppresses low frequency errors in the regime of linear response.

\section{Solving the constraints}
\label{sec.solving-the-constraints}
The precise mathematical problem we attempt to solve is restated as follows. For a given $n$, we need to choose a sequence length $N$ (an odd integer) and seek a sequence $\phi_1, \phi_2, \cdots, \phi_N$ that satisfies Eq.\eqref{eq.overall-constraint} and
\begin{align}\mathbf{c}'_{a, p} = \sum_{l=1}^N(l-1)^p\boldsymbol{\rho}(\phi_l') &= 0 \quad\ \text{for } p = 0, 1, \cdots, n \label{eq.statement-of-problem-2}
\end{align}
where $\phi_l'$ are the toggling-frame phases previously defined, and $\boldsymbol{\rho}(\phi_l') \equiv (\cos\phi_l', \sin\phi_l', 0)$. The $z$-components of Eqs.\eqref{eq.statement-of-problem-2} are always 0, so we can swap the $xy$-plane for the complex plane to make the notation more succinct,
\begin{equation}
	\sum_{l=1}^N(l-1)^p\mathrm{e}^{\mathrm{i}\phi_l'} = 0\quad\ \text{for } p = 0, 1, \cdots, n. \label{eq.constraint-complex}
\end{equation}
We solve for $\phi_l'$ and then use the inverse of Eqs.\eqref{eq.toggling-frame-phases} to get $\phi_l$. Trial and error suggests the general solution of the Eqs.\eqref{eq.constraint-complex} is non-trivial. 
It has been shown that the Weierstrass substitution $\tan(\phi_l/2) = t_l$ converts such transcendental equations into a system of polynomial equations that may be solved with the method of Gröbner basis \cite{low_optimal_2014}.
However, for small $n$ and $N$, elementary algebra is sufficient. For larger $n$ and $N$, numerical solutions may be attempted.

\subsection{Analytically: $n = 1$}
\label{sec.analytically}
When $n = 0$, the constraints only concern suppression of constant errors. To look at time-dependent errors, we start with $n = 1$, \textit{i.e.}, the $\text{PLA}(1)$ criteria, and attempt a solution with $N = 5$. After making linear combinations of Eqs.\eqref{eq.statement-of-problem-2}, we can obtain two new constraints that are symmetric about the center pulse,
\begin{align}
	\mathrm{e}^{\mathrm{i}\phi_1'} + \mathrm{e}^{\mathrm{i}\phi_2'} + 1 + \mathrm{e}^{\mathrm{i}\phi_4'} + \mathrm{e}^{\mathrm{i}\phi_5'} &= 0 \\
	-2\mathrm{e}^{\mathrm{i}\phi_1'} - \mathrm{e}^{\mathrm{i}\phi_2'} + \mathrm{e}^{\mathrm{i}\phi_4'} + 2\mathrm{e}^{\mathrm{i}\phi_5'} &= 0,
\end{align}
where we have set $\phi_3' = 0$ without loss of generality. Limiting the domains of the phases to $-\pi < \phi_l' \leq \pi$ and applying Euler's identity yields
\begin{align}
	1 + 2\mathrm{e}^{\mathrm{i}\alpha}\cos\beta + 2\mathrm{e}^{\mathrm{i}\gamma}\cos\delta &= 0 \label{eq.SC1-eq1}\\
	4\mathrm{i}\mathrm{e}^{\mathrm{i}\alpha}\sin\beta + 2\mathrm{i}\mathrm{e}^{\mathrm{i}\gamma}\sin\delta &= 0, \label{eq.SC1-eq2}
\end{align}
where we define
\begin{align}
	\alpha \equiv \frac{\phi_5' + \phi_1'}{2},\ \beta \equiv \frac{\phi_5' - \phi_1'}{2},\\
	\gamma \equiv \frac{\phi_4' + \phi_2'}{2},\ \delta \equiv \frac{\phi_4' - \phi_2'}{2},
\end{align}
The newly defined angles have ranges $\alpha, \gamma\in(-\pi, \pi]$, and $\beta, \delta\in(-\pi, \pi)$. 

Consider case A: $\sin\beta\neq0$. From Eq.\eqref{eq.SC1-eq2} we have
\begin{equation}
	\mathrm{e}^{\mathrm{i}\alpha} = - \frac{\sin\delta}{2\sin\beta}\mathrm{e}^{\mathrm{i}\gamma}.
\end{equation}
From here we deduce that $\sin\delta = \pm 2\sin\beta \neq 0$ and that $\beta, \delta \neq 0$. Since $\phi_1' = \alpha - \beta \leq \pi$, $\phi_5' = \alpha + \beta \leq \pi$, and $\beta \neq 0$, we must have $\alpha \neq \pi$. Similarly $\gamma \neq \pi$.

One can check that the case $\sin\delta = 2\sin\beta$ leads to a contradiction. Therefore $\sin\delta = -2\sin\beta$. 
Then $\mathrm{e}^{\mathrm{i}\alpha} = \mathrm{e}^{\mathrm{i}\gamma}$, \textit{i.e.}, $\alpha = \gamma$. Plugging this in to Eq.\eqref{eq.SC1-eq1}, we get
\begin{equation}
	1 = -2(\cos\beta + \cos\delta)\mathrm{e}^{\mathrm{i}\alpha}.
\end{equation}
We have that $\alpha = \gamma = 0$, and $\cos\beta + \cos\delta = -\frac{1}{2}$. With trigonometric identities and substitution, we obtain that the solutions to the equations
\begin{align}
	\cos\beta + \cos\delta &= -\frac{1}{2} \\
	\sin\delta &= -2\sin\beta
\end{align}
are $\beta = \pm\arccos\left(\frac{1-2\sqrt{10}}{6}\right)$, $\delta = \mp\arccos\left(\frac{-2+\sqrt{10}}{3}\right)$. Thus,
\begin{align}
	\phi_1' = -\phi_5' = -\beta, \\
	\phi_2' = -\phi_4' = -\delta.
\end{align}
Using Eq.\eqref{eq.toggling-frame-phases}, we convert these phases back to the original frame and add an appropriate phase offset,
\begin{equation}
	\boldsymbol{\phi} = (-\beta, -2\beta+\delta, -2\beta+2\delta, -2\beta+\delta, -\beta).
\end{equation}
We label this solution the $[\text{PLA}(1)]_2$ sequence since the other solution ($[\text{PLA}(1)]_1$, from case B below) is a well known sequence.

Now consider case B: $\sin\beta = 0$. Then from Eq.\eqref{eq.SC1-eq2}, $\beta=\delta=0$. Plugging into Eq.\eqref{eq.SC1-eq1},
\begin{equation}
	1 + 2\mathrm{e}^{\mathrm{i}\alpha} + 2\mathrm{e}^{\mathrm{i}\gamma} = 0.
\end{equation}
Simple geometry yields $\alpha = -\gamma = \pm\arccos\left(-\frac{1}{4}\right)$. So
\begin{align}
	\phi_1' = -\phi_2' = -\phi_4' = \phi_5' = \alpha.
\end{align}
Again using Eq.\eqref{eq.toggling-frame-phases}, we can convert these phases back to the original frame,
\begin{equation}
	\boldsymbol{\phi} = (-3\alpha, -\alpha, 0, \alpha, 3\alpha)  .
\end{equation}
This sequence, which we temporarily denote as $[\text{PLA}(1)]_1$, is in fact the well known $F_1$ sequence discovered by Wimperis \cite{wimperis_iterative_1991, husain_further_2013}. It is time-symmetric in the toggling frame because $\phi_j'=\phi_{6-j}'$. Consequently, $\boldsymbol{\rho}_a(t) = \boldsymbol{\rho}_a(\tau - t)$. If we shift the origin of time to the middle of the sequence, then $\boldsymbol{\rho}_a(t) = \boldsymbol{\rho}_a(-t)$. Incidentally, with this symmetry, the constraints $\mathbf{c}_{a, p}$ evaluate to 0 for all odd power-law drifts,
\begin{equation}
	\mathbf{c}_{a, p} = \frac{1}{p!} \int_{-\tau/2}^{\tau/2} t^p\boldsymbol{\rho}_a(t)\mathrm{d}t = 0 \text{ if }  p \text{ is odd}.
\end{equation}

\subsection{Numerically: $n > 1$}
\label{sec.numerically}
When $n > 1$ and $N$ becomes relatively large, exact solution becomes difficult to obtain analytically, and we solve the constraints Eq.\eqref{eq.overall-constraint} and Eqs.\eqref{eq.statement-of-problem-2} numerically. A simple technique is to define an objective function as the sum of squares of the left hand sides of the constraints and minimize it using a numerical optimization algorithm. Given  $n$, we choose an appropriate sequence length $N$ and define the objective function in terms of the original phases $\phi_l$,
\begin{equation}
	U(\boldsymbol{\phi}) = \sin^2[g(\boldsymbol{\phi})] +\sum_{p=0}^{n}|\mathbf{c}'_{a, p}(\boldsymbol{\phi})|^2.\label{eq:NumericalObjectiveFunction}
\end{equation}
It is a smooth function defined in such a way that it is minimized to 0 when all the constraints are satisfied. Both terms in Eq.~(\ref{eq:NumericalObjectiveFunction}) are non-negative. The first term exploits the periodicity of the sine function and corresponds to constraint Eq.\eqref{eq.overall-constraint} and the second term corresponds to Eqs.\eqref{eq.statement-of-problem-2}. When implementing this objective function in code, it helps to scale each term in $U(\boldsymbol{\phi})$ so that their magnitudes are similar over the domain. A random initial guess may be chosen and a minimization algorithm can be employed to find the optimal $\boldsymbol{\phi}$. If the algorithm terminates at $U(\boldsymbol{\phi}) = 0$, then we have found a solution to the constraints given $n$ for $N$. Otherwise a different initial guess or a larger $N$ may be chosen and the procedure repeated. In general, for each $n$ we are interested in finding the shortest sequence that can satisfy the constraints. When $N$ is too small however, a solution may not be found. Since the numerical procedure previously described does not discern the non-existence of solutions from unfortunate initial guesses, the shortest sequences we discover only provide an upper bound for the sequence length $N$, and even shorter sequences may exist.

\begin{table}[t]
\begin{tabular}{ccc}
\hline
Phases & $\left[\text{PLA}(2)\right]_1$ & $\left[\text{PLA}(3)\right]_1$ \\
\hline
$\phi_1$ & 1.76715945118259 & 4.83865251534654 \\
$\phi_2$ & 5.41431157276639 & 1.84379790507494 \\
$\phi_3$ & 0.60338726707880 & 1.93262975911420 \\
$\phi_4$ & 2.25267362096692 & 0.48888316408261 \\
$\phi_5$ & 5.66568802156378 & 3.13701277837872 \\
$\phi_6$ & 0.11541193070770 & 3.67903366892586 \\
$\phi_7$ & 2.91932560661088 & 3.52519916847217 \\
$\phi_8$ & 3.75846738675240 & 5.73340443857318 \\
$\phi_9$ & 0.58530416475736 & 4.41388024396790 \\
$\phi_{10}$ &  & 4.49690511625724 \\
$\phi_{11}$ &  & 1.53624248122411 \\
\hline
\end{tabular}
\caption{Instances of numerical solutions for $\text{PLA}(2)$ and $\text{PLA}(3)$, with phases given in radians}. Since the numerical procedure that was used to generate theses sequences started with a random initial guess, there can be other solutions that are not listed here.
\label{tab.example-sequence}
\end{table}

\section{Beyond linear response}
\label{sec.beyond-linear-response}
In section \ref{sec.secular-drifts}, we showed that to first order in the Magnus expansion, the sequences we constructed filter out low frequencies. However, the true infidelity (Eq.\eqref{eq.infidelity}) contains higher order terms. We expect that in the low frequency regime, termed the ``DC limit" in \cite{kabytayev_robustness_2014}, $\mathbf{a}_2$ should dominate the infidelity, \textit{i.e.}, $1-\mathcal{F}\approx |\mathbf{a}_2|^2$. At the end of section \ref{sec.secular-drifts} we also drew attention to the potential drawbacks of longer sequence length on infidelity. To examine these effects, we use a Monte Carlo simulation to compute the true infidelity of the composite pulses under random amplitude errors. If we choose the power spectral density of the errors to be narrow-band and centered around $f_c$, then we may scan $f_c$ to numerically obtain the frequency response of the composite pulses. Another important parameter of the power spectral density is the total noise power, or equivalently the RMS error, which affects the relative magnitudes of various order correction terms in the Magnus expansion and causes the breakdown of the linear theory. See appendix \ref{sec.details-about-the-monte-carlo-simulation} for details about the Monte Carlo simulation.

\begin{figure}
  \includegraphics[width=\linewidth]{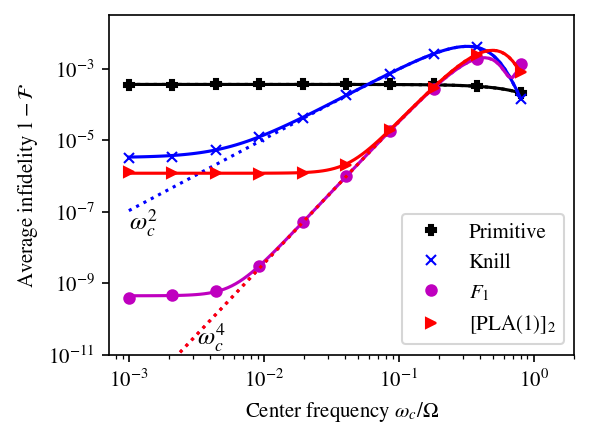}
  \includegraphics[width=\linewidth]{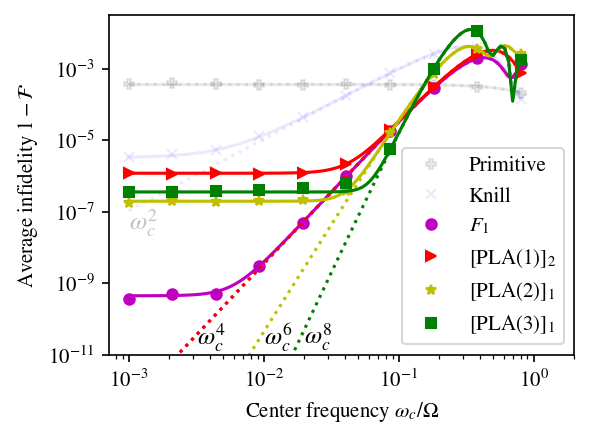}
  \caption{Simulated frequency response of pulse sequences for amplitude errors. The plots are grouped into two figures for legibility and ease of comparison. On the horizontal axis, $\omega_c = 2\pi f_c$ and $\Omega = 1.5\times10^6$ rad/s is the Rabi frequency. The ratio of RMS amplitude error and the Rabi frequency is $1.21\times 10^{-2}$. Markers: average infidelity of a composite pulse obtained by Monte Carlo simulation. Dashed curves: infidelity predicted by the filter function (Eq.\eqref{eq.first-order-theory}). Solid curves: infidelity prediction that includes terms in the Magnus expansion to the third order (Eq.\eqref{eq.first-order-theory} + Eq. \eqref{eq.second-order-theory} + third order correction derived analogously).}
  \label{fig:frequency-scan}
\end{figure}

\begin{figure}
  \includegraphics[width=\linewidth]{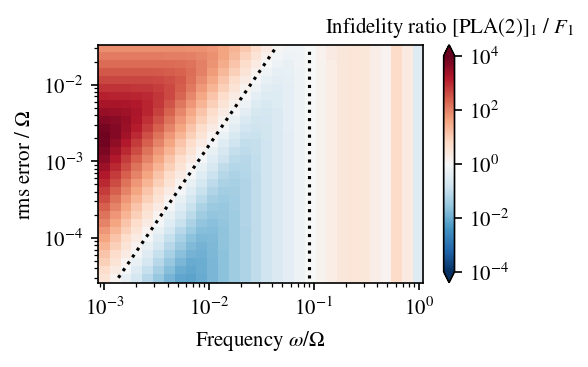}
  \caption{
  Effect of the RMS error on the $[\text{PLA}(2)]_1$ performance bandwidth. Plotted in color is the infidelity ratio between $[\text{PLA}(2)]_1$ and $F_1$,
  calculated using the third-order corrected theory curves (solid curves in figure \ref{fig:frequency-scan}). The blue region indicates the parameters for which $[\text{PLA}(2)]_1$ has a lower infidelity} than $F_1$. A similar shaped blue region exists for $[\text{PLA}(3)]_1$. For the $[\text{PLA}(1)]_2$ sequence however, there is no parameter range over which its infidelity is lower than $F_1$. The dashed lines show the boundaries of the blue region: $\left(\frac{\omega}{\Omega}\right)^2 > \frac{\sigma_{\beta}}{\Omega}(\Omega^3|\mathbf{c}_{a, 2}|)^{-1}\frac{\sqrt{3}}{4}\pi^2\left|\sum_{l=1}^N\sum_{m=1}^{l-1}\sin(\phi_m'- \phi_l')\right|$, and $\frac{\omega}{\Omega} < \frac{|\Omega^3\mathbf{c}_{a, 2}|}{|\Omega^4\mathbf{c}_{a, 3}|}$, where $\sigma_{\beta}$ is the RMS error, $\mathbf{c}_{a, 2}$ is calculated for the $F_1$ sequence, and $\mathbf{c}_{a, 3}$ and $\phi_l'$ is calculated for the $[\text{PLA}(2)]_1$ sequence (See appendix \ref{sec.regime} for derivation of the boundary of regimes).
  \label{fig:noise-power-scan}
\end{figure}

Figure \ref{fig:frequency-scan} shows the simulated frequency response of a number of composite pulses at a fixed RMS error level. Since a composite sequence is to replace a primitive gate as a functional block, we report infidelities without dividing by time or number of pulses. The theory curves in the figure are derived as follows. For narrow-band errors centered around the frequency $f_c$, Eq.\eqref{eq.infidelity-approx} can be approximated as
\begin{align}
	1 - \langle\mathcal{F}\rangle \approx \langle |\mathbf{a}_1|^2\rangle &= \int_{-\infty}^{\infty}\mathrm{d}f\frac{1}{(2\pi f)^2}S_a(f)h_a(f) \\
	&\approx \frac{1}{(2\pi f_c)^2}h_a(f_c)\int_{-\infty}^{\infty}\mathrm{d}fS_a(f), \label{eq.first-order-theory}
\end{align}
where $\int_{-\infty}^{\infty}\mathrm{d}fS_a(f) = \sigma_{\beta}^2$ is the square of the RMS error. To improve the accuracy of the theory curve, we add in the contribution from the second order term in the Magnus expansion. From Eq.\eqref{eq.magnus-expansion-second-order},
\begin{equation}
	\mathbf{a}_2 = \int_0^{\tau}\mathrm{d}t_1\int_0^{t_1}\mathrm{d}t_2 \beta_a(t_1)\beta_a(t_2)\boldsymbol{\rho}_a(t_1)\times\boldsymbol{\rho}_a(t_2).
\end{equation}
This is only important in the low frequency regime, so it suffices to approximate $\beta_a(t) = \beta_a$ and define
\begin{equation}
	\mathbf{a}_{2, \text{DC}} \equiv \beta_a^2\int_0^{\tau}\mathrm{d}t_1\int_0^{t_1}\mathrm{d}t_2 \boldsymbol{\rho}_a(t_1)\times\boldsymbol{\rho}_a(t_2).
\end{equation}
$\boldsymbol{\rho}_a(t)$ is a piece-wise constant function. Substituting it in, we have
\begin{equation}
	\mathbf{a}_{2, \text{DC}} = \frac{1}{4}\left(\frac{\pi}{\Omega}\right)^2\beta_a^2\sum_{l=1}^N\sum_{m=1}^{l-1}\sin(\phi_m'- \phi_l')\mathbf{\hat{z}},
\end{equation}
whose squared average is
\begin{align}
	\langle |\mathbf{a}_{2, \text{DC}}|^2 \rangle &= \frac{1}{16}\left(\frac{\pi}{\Omega}\right)^4\langle \beta_a^4\rangle \left[\sum_{l=1}^N\sum_{m=1}^{l-1}\sin(\phi_m'- \phi_l')\right]^2. \label{eq.second-order-theory}
\end{align}
If $\beta_a$ follows a Gaussian distribution, then $\langle \beta_a^4\rangle = 3\sigma_{\beta}^4$. Therefore we assemble a more accurate prediction of the infidelity by summing the contribution from the filter function and the second order Magnus term. The third order correction can be derived analogously and included as well.

Figure \ref{fig:frequency-scan} shows that the summed theory curves explain the simulated performance of the composite pulses accurately. The Knill sequence, the $F_1$ sequence, and the $[\text{PLA}(1)]_2$ sequence are all composed from five $\pi$ pulses. Each has 4 degrees of freedom after subtracting a global shift of the phases. For the Knill sequence, first-order suppression of constant amplitude and detuning errors establishes two simultaneous constraints on the phases of the pulse \cite{jones_designing_2013, su_quasi-classical_2021}. For $F_1$ and $[\text{PLA}(1)]_2$, the detuning constraint is swapped out for first-order suppression of linearly-varying amplitude errors. The slope of the frequency response shows that, to first order, sequences that satisfy the $\text{PLA}(n)$ criteria become increasingly sharper high-pass filters with increasing $n$. Although the $F_1$ sequence and the $[\text{PLA}(1)]_2$ sequence result from the same set of constraints, the $F_1$ sequence additionally provides second-order suppression of constant amplitude error \cite{wimperis_iterative_1991}. This can also be seen from the fact that it has zero area in the toggling frame (see section V of \cite{su_quasi-classical_2021}). Hence, its DC limit drops more than those of other sequences when the RMS error is reduced.

Since the performance of the sequences depend on the RMS error in addition to the noise frequency, in figure \ref{fig:noise-power-scan} we plot, as an example, the regime over which one might prefer the $[\text{PLA}(2)]_1$ sequence over the $F_1$ sequence on the sole basis of their infidelities, and vice versa. With decreasing RMS error, the contribution from the second order Magnus term drops out from the infidelity, allowing $[\text{PLA}(2)]_1$ to have a lower infidelity over a wider bandwidth. To put some physical sense to this result, let's take the Rabi frequency to be $\Omega = 2 \pi \times 239 \text{ kHz}$. Then the base of the blue triangle in figure \ref{fig:noise-power-scan} spans from approximately $300 \text{ Hz}-20 \text{ kHz}$.  This falls directly in the audio range, where mechanical vibrations in the lab can cause effective amplitude modulation on control tools such as laser beams due to deflection.

\section{Conclusion}
\label{sec.conclusion}
In summary, we have developed the $\text{PLA}(n)$ criteria for designing pulse sequences for single-qubit quantum gates corresponding to $\pi$-rotations about a transverse axis that can suppress, within linear response, errors from pulse amplitude imperfections due to time-dependent secular drifts in experimental instruments. Composite pulses are typically designed for static errors, so they do not satisfy the $\text{PLA}(n)$ criteria. An exception is Wimperis's $F_1$ sequence  \cite{wimperis_iterative_1991}, which obeys the $\text{PLA}(1)$ criteria thanks to its time symmetry. We found analytically that there are only two five-pulse sequences, $[\text{PLA}(1)]_2$ and the $F_1$ sequence, that suppress static error and linear drift to first order in the Magnus expansion. Of the two sequences, $F_1$ is better for suppressing low frequency errors in the limit of small errors because it suppresses static error to second order in the Magnus expansion \cite{wimperis_iterative_1991}. In addition to the $n=1$ case, we found numerically instances of sequences that satisfy the $\text{PLA}(n)$ criteria up to $n = 3$, which suppress power-law drifts up to $t^n$ in the pulse amplitude. These sequences are high-pass filters of filter order $n+1$. While the $F_1$ sequence has a lower DC limit for infidelity, sequences that satisfy the $\text{PLA}(n)$ criteria for $n>1$ have lower infidelities than $F_1$ over a broadening bandwidth as the RMS error decreases. The choice of which sequence to use should include a noise model from the actual system to find the best way to suppress errors, which lends evidence to the need for noise characterization \cite{frey_application_2017, frey_simultaneous_2020}.

The analysis is limited by the assumption of analytical drifts, so it does not extend to jumps or zigzags in the pulse amplitude. Next, since in general one wants to use the shortest possible sequence that suppresses power-law drifts, we are interested in minimizing the sequence length. However, the random initial guess taken by our numerical procedure means that we can not eliminate the existence of shorter sequences than those we found that satisfy the same constraints. In addition, the assumption of Gaussian error distribution means that adjustments are necessary to adopt the analysis for non-Gaussian distributions. Finally, the non-linearity of composite rotations can cause mixing between different frequency components that is not captured by our numerical simulation, where we only inject noise of a very narrow-band.

Future work will investigate how power-law drifts enter higher order terms in the Magnus expansion. Cross-terms between power-law drifts may make it difficult to suppress each drift separately. In addition, we need an estimate for the effect of truncated terms, since they are the ones that eventually remain. Optimal control could be a valuable alternative for designing robust protocol with minimum time or energy. Since it seeks to minimize a cost function, the aforementioned effects due to higher order terms can be naturally taken into account. However, we will need an efficient way to evaluate the cost function which incorporates the average infidelity over the noise ensemble. Another direction for future work is constructing similar sequences for detuning errors, which one might call power-law frequency (PLF) sequences. Suppression of detuning errors has been investigated extensively \cite{ball_role_2016, khodjasteh_designing_2013, cywinski_how_2008, biercuk_dynamical_2011, ball_walsh-synthesized_2015} in the literature for dynamical decoupling. More recently, the case of non-stationary noise was investigated with a generalized notion of filter functions, that are based on the mathematical notion of frames, of which the usual frequency domain representation of the noise spectrum is a special case \cite{chalermpusitarak_frame-based_2021}. Due to the framework's reported capability to handle non-stationary noise, we should investigate the properties of our sequences within this more general framework.

\section{Acknowledgements}
We would like to thank K. R. Brown, M. J. Biercuk, L. Viola, and I. L. Chuang for feedback on the draft and helpful discussions. Work on this project was carried out with support from the Lau family and Ms. Evers-Manly through the UCLA Undergraduate Research Scholars Program, the US National Science Foundation Award No.~PHY-1912555, and the NSF QLCI program through grant number OMA-2016245. Initial work on this project was carried out by Xingchen Fan and Clementine Domine with support from the NSF-DMR under CMMT Grant 1836404.

\appendix

\section{Filter function in the low frequency limit}
\label{sec.filter-function-low-freq}
We will show that if a pulse sequence suppresses power-law drifts, then it will act as a high-pass filter. The definition of the filter function is
\begin{equation}
	h_a(f) = \boldsymbol{\tilde\rho}_a^*(f)\cdot\boldsymbol{\tilde\rho}_a(f),
\end{equation}
where
\begin{equation}
	{\boldsymbol{\tilde\rho}}_a(f)=-\mathrm{i}2\pi f\int_0^{\tau}\boldsymbol{\rho}_a(t)e^{\mathrm{i}2\pi ft}\mathrm{d}t.
\end{equation}
Plugging in gives
\begin{align}
    h_a(f) &= (2\pi f)^2\int_0^{\tau}\int_0^{\tau}\boldsymbol{\rho}_a^*(t_1)\cdot\boldsymbol{\rho}_a(t_2)e^{\mathrm{i}2\pi f(t_2-t_1)}\mathrm{d}t_1\mathrm{d}t_2.
\end{align}
The complex conjugate can be omitted since $\boldsymbol{\rho}_a(t)$ is real. $h_a(f)$ is real valued by definition, so we take its real part,
\begin{align}
    h_a(f) &= (2\pi f)^2\int_0^{\tau}\int_0^{\tau}\boldsymbol{\rho}_a(t_1)\cdot\boldsymbol{\rho}_a(t_2) \nonumber\\
    &\quad\quad \cdot\cos[2\pi f(t_2-t_1)]\mathrm{d}t_1\mathrm{d}t_2 \\
    &= \sum_{n'=0}^{\infty}\frac{(-1)^{n'}}{(2n')!}(2\pi f)^{2n'+2}\int_0^{\tau}\int_0^{\tau}\boldsymbol{\rho}_a(t_1)\cdot\boldsymbol{\rho}_a(t_2) \nonumber\\
    &\quad\quad \cdot(t_2-t_1)^{2n'}\mathrm{d}t_1\mathrm{d}t_2. \label{eq.filter-function-expansion}
\end{align}
The last equality sign invokes the Taylor expansion of cosine. In the low frequency limit ($f\rightarrow0$), the $n'=0$ term dominates, so
\begin{align}
	h_a(f) &\approx (2\pi f)^2\int_0^{\tau}\int_0^{\tau}\boldsymbol{\rho}_a(t_1)\cdot\boldsymbol{\rho}_a(t_2)\mathrm{d}t_1\mathrm{d}t_2 \\
	&=(2\pi f)^2\left(\int_0^{\tau}\boldsymbol{\rho}_a(t)\mathrm{d}t\right)^2.
\end{align}
This term vanishes if $\int_0^{\tau}\boldsymbol{\rho}_a(t)\mathrm{d}t = 0$. The next dominant term scales as $f^4$, making the $h_a(f)$ a better high-pass filter. In general, we apply the binomial expansion to $(t_2-t_1)^{2n'}$ in \eqref{eq.filter-function-expansion},
\begin{align}
    h_a(f) &= \sum_{n'=0}^{\infty}\sum_{m=0}^{2n'}\frac{(-1)^{n'+m}}{{2n'}!}m!(2n'-m)!\begin{pmatrix}2n'\\m\end{pmatrix} \nonumber\\
    &\quad\quad \cdot(\mathbf{c}_{a, m}\cdot\mathbf{c}_{a, 2n'-m})(2\pi f)^{2n'+2},
\end{align}
where, as in the section ``Beyond linear response" in the main text,
\begin{equation}
	\mathbf{c}_{a, p} \equiv \frac{1}{p!}\int_0^{\tau}t^p\boldsymbol{\rho}_a(t)\mathrm{d}t.
\end{equation}
If Eq.(23) in the main text is satisfied, then the limits of summation can be narrowed,
\begin{align}
    h_a(f) &= \sum_{n'=n+1}^{\infty}\sum_{m=n+1}^{2n'-n-1}\frac{(-1)^{n'+m}}{{2n'}!}m!(2n'-m)!\begin{pmatrix}2n'\\m\end{pmatrix} \nonumber\\
    &\quad\quad \cdot(\mathbf{c}_{a, m}\cdot\mathbf{c}_{a, 2n'-m})(2\pi f)^{2n'+2}.
\end{align}
Its leading term is $n' = n + 1$, which simplifies to
\begin{equation}
    h_a(f) \approx \left(\mathbf{c}_{a, n+1}\right)^2\omega^{2n + 4} \label{eq.low-freq-conclusion}
\end{equation}
This quantifies the degree to which composite pulses suppress low frequency noise in the limit that only the first term in the Magnus expansion is needed.

\section{Details about the Monte Carlo simulation}
\label{sec.details-about-the-monte-carlo-simulation}
In the Monte Carlo simulation, we discretize the time domain and integrate the equation of motion of the qubit with the Euler method over an ensemble of numerically generated errors. At the end, we take the ensemble average of the fidelity as defined in Eq.(5). Given a power spectral density, a member of the error ensemble is generated by first dividing the frequency domain into bins ${f_k}$ with width $\Delta f$ and generating one complex Fourier coefficient $A_k$ for each. If the one-sided power spectral density is $S(f)$, then the coefficients are drawn from independent Gaussian distributions,
\begin{align}
	\text{Re}[A_k],\ \text{Im}[A_k] \sim \frac{1}{2}\sqrt{S(f_k)\Delta f} N(0, 1)\quad &\text{ if } f_k \neq 0, \\
	\text{Re}[A_k] \sim \sqrt{S(f_k)\Delta f} N(0, 1),\ \text{Im}[A_k]=0\quad &\text{ if } f_k = 0.
\end{align}
We then take the inverse Fourier transform to obtain the error in the time domain,
\begin{equation}
	\beta(t) = A_0 + \sum_k \left[ A_k \exp\left(\mathrm{i} 2\pi f_k t\right) + \text{c.c.}\right],
\end{equation}
where $A_0$ is the Fourier coefficient for $f=0$, and $\text{c.c}$ denotes the complex conjugate. Consequently, the error has a Gaussian distribution. Here we use the following simple narrow-band (one-sided) power spectral density for the simulation,
\begin{equation}
	S(f) = \begin{cases} A & \text{if } f \in (f_c - f_{\text{band}} / 2, f_c + f_{\text{band}} / 2) \\
	0 & \text{otherwise}\end{cases}
\end{equation}
where $A$ is chosen so that the errors have some known rms value, and $f_{\text{band}}$ is chosen to be 2 Hz, which is sufficiently narrow.

\section{Boundary of regimes}
\label{sec.regime}
In the frequency regime, we can approximate the infidelity of the $F_1$ sequence by
\begin{align}
    1-\langle\mathcal{F}\rangle &\approx \frac{1}{(2\pi f)^2}h_a(f)\sigma_{\beta}^2 \\
    &\approx \frac{1}{(2\pi f)^2}\mathbf{c}_{a, n+1}^2(2\pi f)^{2n+4}\sigma_{\beta}^2,
\end{align}
where $\sigma_{\beta}$ is the RMS error. Since $F_1$ is a solution to the $\text{PLA}(1)$ criteria, we have that $n = 1$. Plugging in,
\begin{align}
    1-\langle\mathcal{F}\rangle &\approx \mathbf{c}_{a, 2}^2(2\pi f)^4\sigma_{\beta}^2
\end{align}
In the low frequency regime, the infidelity of $[\text{PLA}(2)]_1$ is dominated by the second order term in the Magnus expansion. Therefore, assuming Gaussian errors, it infidelity is
\begin{align}
    1 - \langle\mathcal{F}\rangle &\approx \langle |\mathbf{a}_{2, \text{DC}}^2| \rangle \\
    &= \frac{3}{16}\left(\frac{\pi}{\Omega}\right)^4\sigma_{\beta}^4 \left[\sum_{l=1}^N\sum_{m=1}^{l-1}\sin(\phi_m'- \phi_l')\right]^2,
\end{align}
where $\sigma_l'$ are its phases in the toggling frame. At slightly larger frequencies, the infidelity of $[\text{PLA}(2)]_1$ is dominated by the first order term in the magnus expansion. Since for $[\text{PLA}(2)]_1$, we have that $n = 2$. Its infidelity is
\begin{align}
    1 - \langle\mathcal{F}\rangle &\approx \frac{1}{(2\pi f)^2}h_a(f)\sigma_{\beta}^2 \\
    &\approx \frac{1}{(2\pi f)^2}\mathbf{c}_{a, n+1}^2(2\pi f)^{2n+4}\sigma_{\beta}^2 \\
    &= \mathbf{c}_{a, 3}^2(2\pi f)^6\sigma_{\beta}^2
\end{align}
So the lower bound of the blue region is
\begin{align}
    \mathbf{c}_{a, 2}^2(2\pi f)^4\sigma_{\beta}^2 &> \frac{3}{16}\left(\frac{\pi}{\Omega}\right)^4\sigma_{\beta}^4 \left[\sum_{l=1}^N\sum_{m=1}^{l-1}\sin(\phi_m'- \phi_l')\right]^2.
\end{align}
The upper bound is
\begin{align}
    \mathbf{c}_{a, 3}^2(2\pi f)^6\sigma_{\beta}^2 &< \mathbf{c}_{a, 2}^2(2\pi f)^4\sigma_{\beta}^2.
\end{align}
Substituting $\omega = 2\pi f$ and dividing by appropriate factors of $\Omega$,
\begin{align}
    &\left(\frac{\omega}{\Omega}\right)^2 > \frac{\sigma_{\beta}}{\Omega}(\Omega^3|\mathbf{c}_{a, 2}|)^{-1}\frac{\sqrt{3}}{4}\pi^2\left|\sum_{l=1}^N\sum_{m=1}^{l-1}\sin(\phi_m'- \phi_l')\right| \\
    &\frac{\omega}{\Omega} < \frac{|\mathbf{c}_{a, 2}|}{|\Omega\mathbf{c}_{a, 3}|} = \frac{|\Omega^3\mathbf{c}_{a, 2}|}{|\Omega^4\mathbf{c}_{a, 3}|}.
\end{align}
Beware: $\Omega$ and $\sigma_{\beta}$ need to be the same for both sequences. $\mathbf{c}_{a, 2}$ is calculated for to $F_1$. $\phi'$ and $\mathbf{c}_{a, 3}$ are calculated for $[\text{PLA}(2)]_1$.

\section{Non-square pulses}
\label{sec.non-square-pulses}
We will generalize our analysis to non-square pulses. We immediately restrict ourselves to pulse sequences constructed from $N$ equally-spaced $\pi$-pulses, each having duration $t_d$ and whose phases are to be determined. The $l$th pulse begins at time $t_{l-1} = (l-1)t_d$ and ends at time $t_{l}=lt_d$ (denote $\tau \equiv t_N$). The form of the composite pulse sequence is again
\begin{equation}
	\pi_{\phi_1}\to\pi_{\phi_2}\to\cdots\to\pi_{\phi_{N}},
\end{equation}
where the notation $\pi_{\phi_l}$ means a $\pi$-pulse with phase $\phi_l$. In the rotating frame, a pulse $\pi_{\phi_l}$, which accomplishes a rotation of the Bloch vector of the qubit by angle $\pi$ around the axis $\boldsymbol{\rho}_a^{(l)} \equiv \boldsymbol\rho(\phi_l)\equiv(\cos\phi_l,\sin\phi_l,0)$. Assume that the pulses have identical envelopes described by an envelope function $G(t)$ which is non-zero only when $0 < t < t_d$. The instantaneous Rabi frequency during the $l$th pulse is 
\begin{equation}
    \Omega_l(t) = G(t-t_{l-1})(\Omega_0 + \beta_a(t)),
\end{equation}
where $\Omega_0$ is the nominal Rabi frequency and $\beta_a(t)$ models the amplitude error. $G(t)$ is normalized to ensure that the time integral of $\Omega(t)$ for each pulse is $\pi$. In the rotating frame, after suitable approximations, the error-free Hamiltonian ($\hbar \equiv 1$) is given by
\begin{equation}
	H_0(t) = \sum_{l=1}^NG(t-t_{l-1})\frac{\Omega_0}{2}\boldsymbol{\rho}_a^{(l)}\cdot\boldsymbol{\hat\sigma},
\end{equation}
where we define the operator $\boldsymbol{\hat\sigma} \equiv (\hat{\sigma}_x, \hat{\sigma}_y,\hat{\sigma}_z)$. Denote $U_0(t_2, t_1)$ as the unitary operator that evolves the state from time $t_1$ to $t_2$ according to $H_0(t)$ (\textit{i.e.}, when errors are zero). The full Hamiltonian in the rotating frame after suitable approximations is
\begin{align}
	H(t) &= \sum_{l=1}^NG(t-t_{l-1})\frac{[\Omega_0+\beta_a(t)]}{2}\boldsymbol{\rho}_a^{(l)}\cdot\boldsymbol{\hat\sigma} \\
	&= H_0(t) + \beta_a(t)\frac{1}{2}\sum_{l=1}^NG(t-t_{l-1})\boldsymbol{\rho}_a^{(l)}\cdot\boldsymbol{\hat\sigma} \\
	&= H_0(t) + H_{\mathrm{err}}(t).
\end{align}
Now we work in the toggling frame where the effective Hamiltonian is
\begin{equation}
    H'(t) = U_0^{\dagger}(t, 0)H_{\mathrm{err}}(t)U_0(t, 0).
\end{equation}
Suppose $t_{l-1} < t < t_l$, and denote
\begin{align}
    V = U_0(t_{l-1}, 0).
\end{align}
Then
\begin{align}
    H'(t) &= V^{\dagger}U_0^{\dagger}(t, t_{l-1})H_{\mathrm{err}}(t)U_0(t, t_{l-1})V.
\end{align}
where
\begin{align}
    U_0(t, t_{l-1}) = \exp\left[-\mathrm{i}\int_{t_{l-1}}^t\mathrm{d}t'\ G(t'-t_{l-1})\frac{\Omega_0}{2}\boldsymbol{\rho}_a^{(l)}\cdot\boldsymbol{\hat{\sigma}}\right].
\end{align}
Since $H_{\mathrm{err}}(t)$ commutes with $U_0(t, t_{l-1})$, we can write
\begin{align}
    H'(t) &= V^{\dagger}H_{\mathrm{err}}(t)V \\
    &= \frac{1}{2}G(t-t_{l-1})\beta_a(t)V^{\dagger}\boldsymbol{\rho}_a^{(l)}\cdot\boldsymbol{\hat\sigma}V.
\end{align}
As before the operator sandwich evaluates to
\begin{align}
    V^{\dagger}\boldsymbol{\rho}_a^{(l)}\cdot\boldsymbol{\hat\sigma}V = \boldsymbol{\rho}(\phi_l')\cdot\boldsymbol{\hat\sigma},
\end{align}
where $\phi_l'$ are the toggling frame phases, so
\begin{align}
    H'(t) = \frac{1}{2}G(t-t_{l-1})\beta_a(t)\boldsymbol{\rho}(\phi_l')\cdot\boldsymbol{\hat\sigma}.
\end{align}
Now let $t$ range from $0$ to $\tau$. Then,
\begin{align}
    H'(t) = \frac{1}{2}\sum_{l=1}^NG(t-t_{l-1})\beta_a(t)\boldsymbol{\rho}(\phi_l')\cdot\boldsymbol{\hat\sigma}.
\end{align}
The next step is a key definition that is different from the main text. Define
\begin{align}
    \boldsymbol{\rho}_a(t) \equiv \frac{1}{2}\sum_{l=1}^NG(t-t_{l-1})\boldsymbol{\rho}(\phi_l').
\end{align}
Then the toggling frame Hamiltonian reduces to the envelope independent form
\begin{align}
    H'(t) = \beta_a(t)\boldsymbol{\rho}_a(t)\cdot\boldsymbol{\hat\sigma}.
\end{align}
If we take $t_d \to \pi / \Omega_0 $ and $G(t) \to 1$ whenever nonzero, then we recover the case discussed in the main text. Effectively the using non-square pulses introduces a multiplicative periodic sampling function in the definition of $\boldsymbol{\rho}_a(t)$. This means that any analysis we did in the main text that does not invoke explicit definition of $\boldsymbol{\rho}_a(t)$ will remain sound. Now we examine the consequence of this substitution on various quantities whose derivation depends on the explicit form of $\boldsymbol{\rho}_a(t)$.

\subsection{The constraints $\mathbf{c}_{a, p}$ and $\mathbf{c}'_{a, p}$}
$\mathbf{c}_{a, p}$ is the constraint for suppressing the $p$th order power drift. It is defined as
\begin{equation}
	\mathbf{c}_{a, p} \equiv \frac{1}{p!}\int_0^{\tau}t^p\boldsymbol{\rho}_a(t)\mathrm{d}t,
\end{equation}
Substituting in $\boldsymbol{\rho}_a(t)$ yields
\begin{align}
	\mathbf{c}_{a, p} &= \frac{1}{2p!}\int_0^{\tau}t^p\sum_{l=1}^NG(t-t_{l-1})\boldsymbol{\rho}(\phi_l')\mathrm{d}t \\
	&= \frac{1}{2p!}\sum_{l=1}^N\boldsymbol{\rho}(\phi_l')\int_{-\infty}^{+\infty}G(t-t_{l-1})t^p\mathrm{d}t \\
	&= \frac{1}{2p!}\sum_{l=1}^N\boldsymbol{\rho}(\phi_l')\int_{-\infty}^{+\infty}G(t)(t+t_{l-1})^p\mathrm{d}t \\
	&= \frac{1}{2}\sum_{l=1}^N\boldsymbol{\rho}(\phi_l')\sum_{q=0}^p\frac{1}{q!(p-q)!}(t_{l-1})^q\int_{-\infty}^{+\infty}G(t)t^{p-q}\mathrm{d}t.
\end{align}
where we again invoke the binomial theorem in the last step. Define the non-zero integrals
\begin{align}
    I_{p-q} \equiv \int_{-\infty}^{+\infty}G(t)t^{p-q}\mathrm{d}t.
\end{align}
Then
\begin{align}
	\mathbf{c}_{a, p} &= \frac{1}{2}\sum_{q=0}^p\sum_{l=1}^N\frac{I_{p-q}}{q!(p-q)!}(t_{l-1})^q\boldsymbol{\rho}(\phi_l') \\
	&= \frac{1}{2}\sum_{q=0}^p\frac{I_{p-q}t_d^q}{q!(p-q)!}\sum_{l=1}^N(l-1)^q\boldsymbol{\rho}(\phi_l') \\
	&= \frac{1}{2}\sum_{q=0}^p\frac{I_{p-q}t_d^q}{q!(p-q)!}\mathbf{c}'_{a, q},
\end{align}
where again we define
\begin{equation}
	\mathbf{c}'_{a, p} \equiv \sum_{l=1}^N(l-1)^p\boldsymbol{\rho}(\phi_l').
\end{equation}
Therefore the $\text{PLA}(n)$ criteria are unchanged. However, the exact numerical values of $\mathbf{c}_{a, p}$ which determine the numerical value of the infidelity, will be different.

\subsection{The filter function $h_a(f)$}
The definition of the filter function depends on $\boldsymbol{\rho}_a(t)$, but the derivation in section \ref{sec.filter-function-low-freq} does not invoke the explicit form of $\boldsymbol{\rho}_a(t)$. Therefore the conclusion, Eq.\eqref{eq.low-freq-conclusion}, still holds. The first order frequency response in the low-frequency regime will be shifted up or down because of the modified numerical value of $\mathrm{c}_{a, n+1}$

\subsection{Beyond linear response: $\mathbf{a}_{2, DC}$}
In the main text, $\mathbf{a}_{2, DC}$ quantifies the second-order correction in the Magnus expansion for very low frequencies. It is defined as
\begin{equation}
	\mathbf{a}_{2, \text{DC}} \equiv \beta_a^2\int_0^{\tau}\mathrm{d}t_1\int_0^{t_1}\mathrm{d}t_2 \boldsymbol{\rho}_a(t_1)\times\boldsymbol{\rho}_a(t_2).
\end{equation}
$\boldsymbol{\rho}_a(t)$ is no longer a piece-wise constant function, but it almost behaves like one. Substituting it in, we have
\begin{align}
	\mathbf{a}_{2, \text{DC}} &= \frac{\beta_a^2}{4}\sum_{l,m=1}^N\boldsymbol{\rho}(\phi_l')\times\boldsymbol{\rho}(\phi_m') \nonumber \\
	&\quad\quad\cdot \int\displaylimits_{t_2<t_1}\mathrm{d}t_1\mathrm{d}t_2 G(t_1-t_{l-1})G(t_2-t_{m-1}).
\end{align}
The integral is non-zero only if $m \leq l$, but the cross product evaluates to zero for $m=l$, so
\begin{align}
	\mathbf{a}_{2, \text{DC}} &= \frac{\beta_a^2}{4}\sum_{m < l}^N\boldsymbol{\rho}(\phi_l')\times\boldsymbol{\rho}(\phi_m') \nonumber \\
	&\quad\quad\cdot \int\displaylimits_{t_2<t_1}\mathrm{d}t_1\mathrm{d}t_2 G(t_1-t_{l-1})G(t_2-t_{m-1}).
\end{align}
$G(t_1-t_{l-1})$ is nonzero only if $t_{l-1}<t_1<t_l$. When $m < l$, we have that $t_m \leq t_{l-1}$. For all $t_2 > t_1$ and $t_{l-1}<t_1<t_l$, we must have $t_m \leq t_{l-1} < t_1 <t_2$, so $G(t_2-t_{m-1})$ must be zero. Therefore the integrand is 0 for $t_2 > t_1$, and we can extend the limits of integration,
\begin{align}
	\mathbf{a}_{2, \text{DC}} &= \frac{\beta_a^2}{4}\sum_{m < l}^N\boldsymbol{\rho}(\phi_l')\times\boldsymbol{\rho}(\phi_m') \nonumber \\
	&\quad\quad\cdot \int\int\mathrm{d}t_1\mathrm{d}t_2 G(t_1-t_{l-1})G(t_2-t_{m-1}). \\
	&= \frac{\beta_a^2}{4}\left(\int\mathrm{d}t_1 G(t_1)\right)^2\sum_{m < l}^N\boldsymbol{\rho}(\phi_l')\times\boldsymbol{\rho}(\phi_m') \\
	&= \frac{1}{4}I_0^2\beta_a^2\sum_{m < l}^N\sin(\phi_m'- \phi_l')\mathbf{\hat{z}}.
\end{align}
This result differs from that in the main text only by a constant factor. $\mathbf{a}_{3, DC}$ can be similarly derived. The combined effect on the conclusions of in the main text is that the DC limit will be shifted by some constant factor, and the boundaries between different regimes over which one might prefer one sequence over another on the basis of infidelity may shift.

\bibliography{References.bib}

\end{document}